\documentclass[a4paper]{article}

\usepackage{INTERSPEECH2018}
\usepackage{graphicx}
\usepackage{booktabs}
\usepackage{multirow}
\usepackage{xcolor}

\title{Towards Learning Fine-Grained Disentangled Representations from Speech}

\name{Yuan Gong, Christian Poellabauer}

\address{
Department of Computer Science and Engineering\\
University of Notre Dame, IN 46556, USA
}
\email{ygong1@nd.edu, cpoellab@nd.edu}

\begin{document}

\maketitle

\begin{abstract}
Learning disentangled representations of high-dimensional data is currently an active research area. However, compared to the field of computer vision, less work has been done for speech processing. In this paper, we provide a review of two representative efforts on this topic and propose the novel concept of fine-grained disentangled speech representation learning.

\end{abstract}

\section{Introduction}

Representation learning is a fundamental challenge in machine learning and artificial intelligence. While there are multiple criteria for an ideal representation, \textbf{disentangled  representation} (illustrated in Figure 1), which explicitly separates the underlying causal factors of the observed data, has been of particular interest, because it can be useful for a large variety of tasks and domains~\cite{300higgins2016beta,301chen2016infogan,302kulkarni2015deep,303banijamali2017jade,304reed2014learning}. For example, in~\cite{304reed2014learning}, the authors show that learning disentangling latent factors corresponding to pose and identity in photos of human faces can improve the performance of both pose estimation and face verification. Learning disentangled representation from high-dimensional data is not a trivial task and multiple techniques, such as $\beta$-VAE~\cite{300higgins2016beta}, InfoGAN~\cite{301chen2016infogan}, and DC-IGN~\cite{302kulkarni2015deep}, have been developed to address this problem. While disentangling natural image representation has been  studied extensively, much less work has focused on natural speech, leaving a rather large void in the understanding of this problem. In this paper, we first present a short review and comparison of two representative efforts on this topic~\cite{205hsu2017unsupervised,201chou2018multi}, where both efforts involve using an auto-encoder and can be applied to the same task (i.e., voice conversion), but the key disentangling algorithms and underlying ideas are very different.

In~\cite{205hsu2017unsupervised}, the authors proposed an unsupervised factorized hierarchical variational autoencoder (FHVAE). The key idea is that assuming that the speech data is generated from two separate latent variable sets $\boldsymbol{z_1}$ and $\boldsymbol{z_2}$, where $\boldsymbol{z_1}$ contains segment-level (short-term) variables and $\boldsymbol{z_2}$ contains sequence-level (long-term) variables ($\boldsymbol{z_2}$ that are further conditioned on an s-vector $\boldsymbol{\mu_2}$). Leveraging the \emph{multi-scale nature} that different factors affect speech at different time scales (e.g., speaker identity affects the fundamental frequency and volume of speech signal at the sequence level while the phonetic content affects the speech signal at the segment level), by training an autoencoder in a sequence-to-sequence manner, $\boldsymbol{z_1}$ can be forced to encode segment-level information (e.g., speech content), while $\boldsymbol{z_2}$ and $\boldsymbol{\mu_2}$ can be forced to encode sequence-level information (e.g., speaker identity). In the experiments, by keeping  $\boldsymbol{z_2}$ fixed and changing $\boldsymbol{z_1}$, speech of the same content, but by different speakers, can be synthesized naturally, demonstrating the clean separation between content and speaker information. Further, the learned s-vector is shown to be a stronger feature than the conventional i-vector in the speaker verification task, demonstrating that it encodes speaker-level information well. In~\cite{205hsu2017unsupervised} and subsequent efforts~\cite{203hsu2018unsupervised,208hsu2018extracting, tang2018study}, the authors further showed that the disentangled representation is also helpful in the speech recognition task. These efforts  convey two primary insights: 1) by adding the appropriate prior assumptions on the latent variables, speech content information and speaker-level information can be separated out in an unsupervised learning manner; 2) the learned disentangled representations are useful to improve both speech synthesis and broader inference tasks.

\begin{figure}[t]
  \center
  \includegraphics[width=3.3cm]{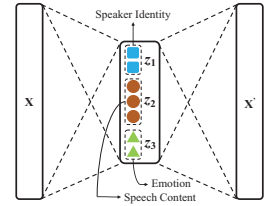}
  \caption{Illustration of disentangled representations (using an auto-encoder as an example). We expect to learn a structured latent variable set $\boldsymbol{Z}$=$\{\boldsymbol{z_1},...,\boldsymbol{z_n}\}$, where each element corresponds to one independent underlying causal factor of the data.}
  \label{fig:app}
\end{figure}

Different from~\cite{205hsu2017unsupervised}, in~\cite{201chou2018multi}, the authors propose a supervised approach based on adversarial training~\cite{71lample2017fader,72ganin2016domain,73louppe2017learning,choi2017stargan}(illustrated in Figure~\ref{fig:model} (left)). In addition to a regular auto-encoder, the authors add a regularization term in its objective function to force the latent variables (i.e., the encoding) to not contain speaker information. This is done by introducing an auxiliary speaker verification classifier $\boldsymbol{C}$. $\boldsymbol{C}$ is trained to correctly identify the speaker $y$ from the latent variables $\boldsymbol{z}$ (i.e., minimizing the misclassification loss $L_c=-logP(y|\boldsymbol{z})$), while the encoder is trained to maximize $L_c$, i.e., to avoid encoding speaker information in $\boldsymbol{z}$. Both $\boldsymbol{z}$ and speaker label $y$ are fed to the decoder for reconstruction, and the complete objective function of the auto-encoder is hence minimizing $L_{rec} -\lambda L_c$ (where $L_{rec}$ is the point-wise L1-norm loss). By alternatively training the auto-encoder and $\boldsymbol{C}$, the $\boldsymbol{z}$ is learned to be an encoding of speech content information. Further, the residual information of the speech is reconstructed through another GAN and auxiliary classifier. The experiment shows that such a scheme can be successfully applied to a voice conversion task. The main insight of this work is how to use supervision to conduct representation disentanglement.

In summary, although the implementation is very different, in order to learn disentangled representations, both works add constraints to the latent variables. Such a constraint can be a prior assumption (in the unsupervised case) or a regularization term (in the supervised case). While both efforts show good empirical performance in real tasks and lay the groundwork for future efforts, the learned disentangled representation is relatively \textbf{coarse-grained}. That is, in~\cite{205hsu2017unsupervised}, $\boldsymbol{z_1}$ and $\boldsymbol{z_2}$ are in fact corresponding to general fast-changing and slow-changing information, i.e., $\boldsymbol{z_1}$ may contain other fast-changing information such as emotion, while $\boldsymbol{z_2}$ may contain slow-changing factors such as background and channel noise. In~\cite{201chou2018multi}, the authors actually separate out speaker information and general non-speaker information, which may contain a lot of detailed information. Coarse-grained disentangled representations are enough for some tasks, such as voice speaker conversion, but might be limited for other tasks. Next, we discuss the need for (and benefits of) fine-grained disentangled representations. 

\section{Fine-Grained Disentangled Speech Representation}

Natural speech signal inherently contains both linguistic and paralinguistic information; common paralinguistic information include gender, age, health status, personality, friendliness, mood, and emotion (sorted from long-term to short-term)~\cite{210schuller2013computational}.  In the original representation of natural speech, these information are entangled together. But in fact, many of these information are essentially independent or of low correlation with each other as well as with the linguistic content, hence raising the possibility to disentangle them in some latent space. Natural speech signals can be viewed as produced by multiple fine-grained causal factors and disentangling these factors leads to the following benefits:

\textbf{Synthesis:} Learning fine-grained disentangled representation can help more flexible speech synthesis. Assume that disentangled latent variables corresponding to age, personality, friendliness, emotion, and content are learned; we may then be able to synthesize speech signals corresponding to arbitrary combinations of these factors, according to the requirement of the application scenario. Further, this may support novel AI applications, such as speech style transfer and predicting the future voice of a given subject (similar technology has been adopted in computer vision, e.g., image style transfer~\cite{601gatys2016image} and face aging~\cite{602antipov2017face}). In contrast, a coarse-grained disentangled representation~\cite{205hsu2017unsupervised,201chou2018multi} may only support a simple voice speaker conversion task. 

\textbf{Inference:} Learning fine-grained disentangled representation can also help with more accurate inference and reasoning. When we attempt to predict one target variable, we usually want to eliminate the interference of other factors. For example, a speech recognition system is expected to be emotion-independent, while a speech emotion recognition system is expected to be text-independent. Historically, some manually designed algorithms are used to eliminate the effects of unrelated factors, e.g., \emph{speaker normalization}~\cite{501sethu2007speaker,502busso2011iterative} and \emph{speaker adaptation}~\cite{503ding2012speaker,504rahman2012personalized} are commonly used to eliminate the impacts of speaker variability. However, it is difficult to manually design algorithms for all underlying factors. Previous work in representation learning has shown that by disentangling different (independent) factors, all corresponding inference tasks~\cite{304reed2014learning} can gain performance improvements. In addition, the learned disentangled and interpretable representation helps us to understand the inner working mechanism of the machine learning model.

The question is how can we learn fine-grained disentangled representation from speech? Following the discussion in the previous section, we can either add a prior assumption (unsupervised) or a regularization (supervised) to the latent variables. However, for unsupervised approaches, when we want to disentangle many factors, designing such prior assumptions is difficult and needs to be done very carefully. Hence, we first consider a fully supervised solution.

\begin{figure}[t]
  \centering
  \includegraphics[width=8cm]{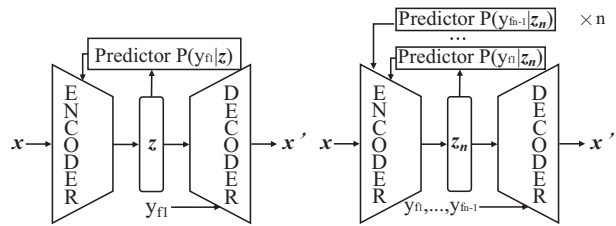}
  \caption{Illustration of supervised representation learning. Left: the approach in~\cite{201chou2018multi}. Right: the proposed approach for learning fine-grained disentangled representation.}
  \label{fig:model}
\end{figure}

Assume that we want to disentangle $n$ independent factors $f_1$, $f_2$, ...,$f_n$ of natural speech. Further assume that we have a data set $\boldsymbol{D}$ that has complete annotations \{$y_{f_{1}}$,$y_{f_{2}}$,...,$y_{f_{n}}$\} corresponding to each factor for each sample. We can then extend the approach in~\cite{201chou2018multi} to learn disentangled latent variables \{$\boldsymbol{z_{f_{1}}}$,$\boldsymbol{z_{f_{2}}}$,...,$\boldsymbol{z_{f_{n}}}$\} corresponding to more than two factors. As illustrated in Figure~\ref{fig:model} (right), we build $n$ auto-encoders, each used to learn latent variables corresponding to one factor. In order to guarantee the disentanglement, for each auto-encoder we further need to build $n-1$ auxiliary predictors. Since the auto-encoder attempts to learn latent variable $\boldsymbol{z_{f_{i}}}$, we train each predictor to correctly predict one factor $j (j\ne i)$ based on $\boldsymbol{z_{f_{i}}}$ (i.e., minimizing the miss prediction loss $L_{ij}$), and train the auto-encoder to maximize the minimum of the loss of each predictor. The training is conducted alternatively and during the training, ground truth annotations $\mathbf{y}=\{y_j|j\ne i\}$ are fed to the decoder for successful reconstruction. Hence, the complete loss of auto-encoder $i$ is $L_i = L_{rec} - \min\limits_{j\ne i}(\lambda_jL_{ij})$, where $L_{rec}$ is the point-wise L1-norm reconstruction loss, and $\lambda$ is the parameter controlling the disentanglement degree of each factor. Note that the predictors having the same target factor cannot be re-used across different auto-encoders, because they are based on different latent variables.

However, in practice, there does not exist such a speech dataset $\boldsymbol{D}$ that has complete annotations for each factor; most speech datasets only have a limited number of annotations. Nevertheless, natural speech has relatively fixed variation factors, which makes it possible to use multiple datasets to cooperatively learn disentangled representations. This is very different from natural images, which have a much larger variation. For example, handwriting digit images have factors ``number'' and ``writing style'', which are completely different from human face images; therefore, it is difficult to use a handwriting image dataset and a human face dataset together to learn disentangled representations. In contrast, for most speech datasets, although they are collected for different purposes and hence have different annotations, they share the same factors such as content, emotion, age, and gender.

Assume that we have a set of datasets $\boldsymbol{D_1}$, $\boldsymbol{D_2}$, ...,$\boldsymbol{D_n}$, where each has an annotation of one different factor. In this practical setting, two things become more complicated: 1) without the ground-truth label, the loss of predictor $\log P(y|\boldsymbol{z})$ cannot be calculated; 2) the encoder aims to remove the information unrelated to the desired factor and the decoder does not have the ground-truth label of the removed information to reconstruct the input. To solve these challenges, we only use the samples with the corresponding ground-truth labels to train the predictors, and use the certainty of the prediction as the regularizer for the auto-encoder (e.g., for discrete labels $\mathbf{y}$: $\max P(y_k|\boldsymbol{z}), k\in{[1,Card(\mathbf{y})]}$). In the same iteration, we also feed this prediction to the decoder. In this approach, we penalize the high-certainty prediction rather than the correct prediction, hence the ground-truth label is not needed. We feed the prediction as the ground truth to the decoder, compensating the information the encoder removes (according to this prediction) to reconstruct the input. 

In this paper, we extend the idea of~\cite{201chou2018multi} for fine-grained disentangled speech representation learning. The training procedure detail, convergence condition, and the empirical performance need to be further explored and are left as future work.

\bibliographystyle{IEEEtran}

\bibliography{reference}

\end{document}